\documentclass[
 reprint,
 amsmath,amssymb,
 aps,
 prl,
]{revtex4-1}

\usepackage{graphicx}
\usepackage{dcolumn}
\usepackage{bm}
\usepackage{color}

\begin{document}

\title{Composite Fermions with Tunable Fermi Contour Anisotropy}

\author{D.\ Kamburov}
\author{Yang\ Liu}
\author{M.\ Shayegan}
\author{L.N.\ Pfeiffer}
\author{K.W.\ West}
\author{K.W.\ Baldwin}
\affiliation{Department of Electrical Engineering, Princeton University, Princeton, New Jersey 08544, USA}

\date{\today}

\begin{abstract}

The composite fermion formalism elegantly describes some of the most fascinating behaviours of interacting two-dimensional carriers at low temperatures and in strong perpendicular magnetic fields. In this framework, carriers minimize their energy by attaching two flux quanta and forming new quasi-particles, the so-called composite fermions. Thanks to the flux attachment, when a Landau level is half-filled, the composite fermions feel a vanishing effective magnetic field and possess a Fermi surface with a well-defined Fermi contour. Our measurements in a high-quality two-dimensional hole system confined to a GaAs quantum well demonstrate that a parallel magnetic field can significantly distort the hole-flux composite fermion Fermi contour.

\end{abstract}


\maketitle

High-quality two-dimensional (2D) carrier systems offer rich opportunities for exploring new physical phenomena. At very low temperatures and in the presence of a strong perpendicular magnetic field ($B_{\perp}$), the electron-electron interaction in these systems leads to a variety of remarkable many-body phases, examples of which include the fractional quantum Hall effect (FQHE) state, the Wigner crystal, and the non-uniform density phases such as stripe and bubble phases \cite{DasSarma.1997,Shayegan.2006,Jain.2007}. The FQHE can be successfully described through the concept of composite fermions (CFs), quasi-particles formed by the attachment of two (or in general an even number of) flux quanta to each carrier in high $B_{\perp}$ \cite{Jain.2007,Jain.PRL.1989,Halperin.PRB.1993,Willett.PRL.1993,Kang.PRL.1993,Goldman.PRL.1994,Smet.PRL.1996}. At the applied magnetic field $B_{\perp,1/2}$ where the lowest Landau level is exactly half-filled ($\nu=1/2$), the flux attachment completely cancels this external field, leaving the CFs as if they are at zero \textit{effective} magnetic field. The effective field the CFs feel away from $\nu=1/2$ is given by $B^*_{\perp}=B_{\perp}-B_{\perp,1/2}$ \cite{footnote100}. At and near $\nu=1/2$, analogously to the low-field carriers, the CFs occupy a Fermi sea with a well-defined Fermi contour.

The existence of a CF Fermi contour raises the question whether any low-field Fermi contour anisotropy is transmitted to the high-field CFs after fermionization \cite{Balagurov.PRB.2000,Gokmen.Nature.2010}. This issue was partially addressed in a recent experimental study of 2D electrons confined to an AlAs quantum well where they have an anisotropic (elliptical) Fermi contour \cite{Gokmen.Nature.2010}. The study revealed that, qualitatively similar to their $B_{\perp}=0$ electron counterparts, CFs also exhibit a \textit{transport} anisotropy. Namely, the resistance at $\nu=1/2$ is larger along the long axis of the $B_{\perp}=0$ electron Fermi contour (where the effective mass is large) compared to the resistance along the short axis (where the effective mass is smaller). While this observation suggests that the CFs might also possess an anisotropic Fermi contour, it does not provide conclusive or quantitative evidence for such anisotropy. An anisotropy in the CF scattering time, for example, would also lead to anisotropic transport. More generally, the problem of anisotropy in FQHE phenomena has sparked recent interest both experimentally and theoretically \cite{Xia.Nat.Phys.2011, Mulligan.PRB.2010, Yang.Haldane.PRB.2012, Qui.Haldane.PRB.2012, Wang.PRB.2012}. Here we report direct measurements evincing that the CF Fermi contour can be anisotropic. Moreover, we demonstrate that the anisotropy is tunable via the application of a strong magnetic field parallel to the 2D plane.

\begin{figure}[b!]
\includegraphics[trim=0cm 0cm 0cm 0cm, clip=false, width=.49\textwidth]{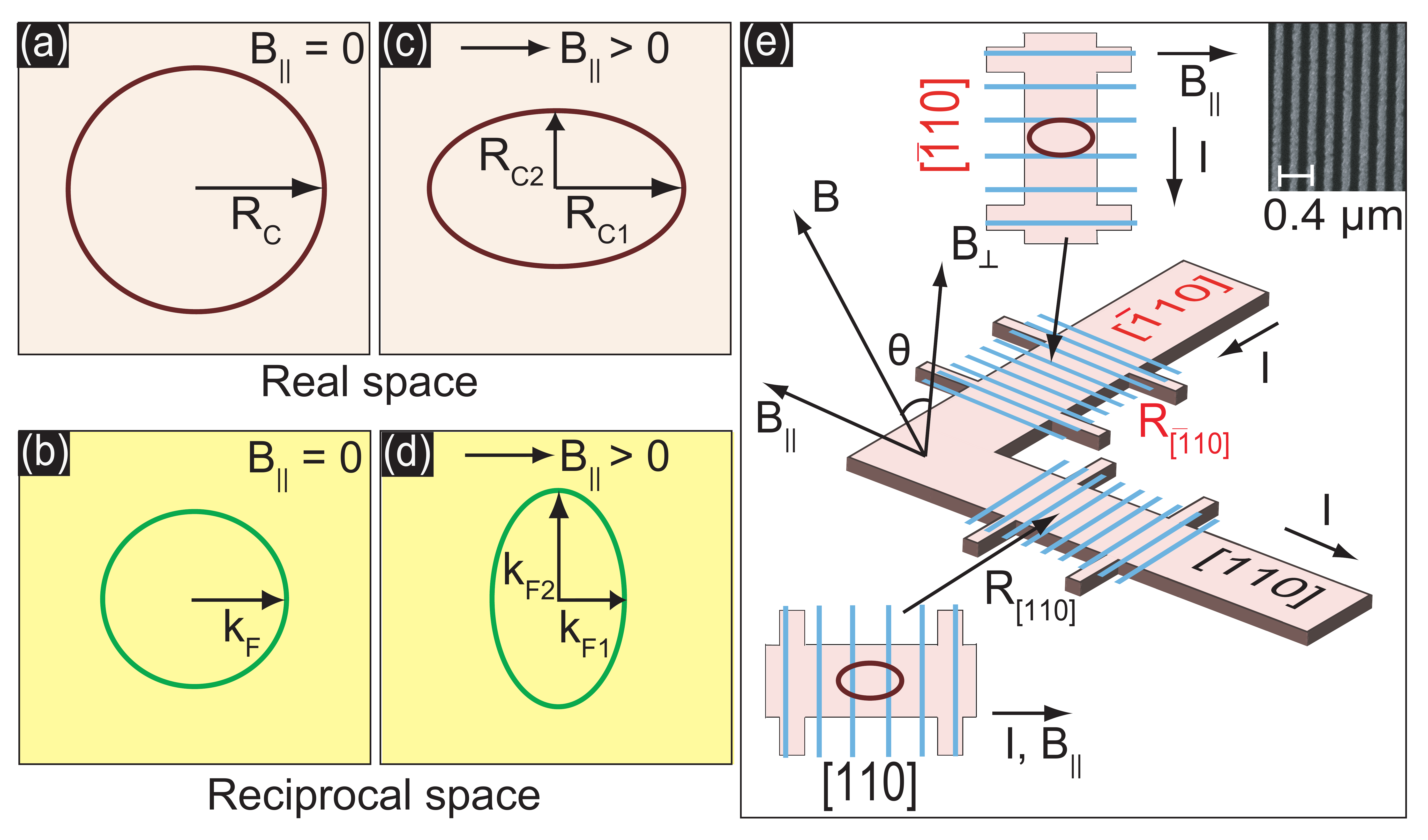}
\caption{\label{fig:Fig1} (color online) (a) and (b) The cyclotron orbit and the Fermi contour are shown, respectively, for an isotropic 2D system when $B_{||}=0$. (c), (d) If the 2D system has a finite (non-zero) thickness, applying $B_{||}>0$ distorts the cyclotron orbit and the Fermi contour. (e) The sample has two Hall bars along the perpendicular directions [$110$] and [$\overline{1}10$], and $B_{||}$ is introduced along the [$110$] direction by tilting the sample with respect to the the magnetic field direction. The electron-beam resist grating covering the top surface of each Hall bar is shown as blue stripes. The orientations of the Hall bars and the resist gratings are chosen to probe the Fermi contours in the [$110$] and [$\overline{1}10$] directions. The cyclotron orbits, given with brown lines, are shown for the case when the orbit diameter fits the grating period $a$ in the [$\overline{1}10$] direction but is larger than $a$ in the [110] direction. Inset: Scanning electron microscope image of the electron-beam resist grating with an $a=200$ nm period. }
\end{figure}

Figure\;\ref{fig:Fig1} highlights the main ingredients of our study. Imagine an isotropic 2D system in which the charged particles have a circular Fermi contour (in reciprocal space) with Fermi wave vector $k_F$ (Fig. 1(b)). In a small, purely perpendicular, magnetic field the particles' classical cyclotron orbit is also circular and is completely characterized by the cyclotron radius $R_{C}$ (Fig. 1(a)). Now, if the particles have a finite (non-zero) layer thickness, a parallel magnetic field ($B_{||}$) applied in the 2D plane couples to their out-of-plane orbital motion and leads to a deformation of the cyclotron orbit, shrinking its diameter in the in-plane direction perpendicular to $B_{||}$ (Fig. 1(c)). Equivalently, the particles' Fermi contour becomes elongated in the direction perpendicular to $B_{||}$ (Fig.\;\ref{fig:Fig1}(d)).

The Fermi contour and/or the cyclotron orbit deformation can be directly probed in a sample with a small, periodic, one-dimensional, density modulation where the carriers complete ballistic cyclotron orbits: whenever the orbit diameter becomes commensurate with the period of the density modulation, the sample's magneto-resistance exhibits a resistance minimum. In particular, the anisotropy of the cyclotron orbit or the Fermi contour can be determined via measuring the positions of the commensurability magneto-resistance minima along the two perpendicular arms of an L-shaped Hall bar as shown in Fig.\;\ref{fig:Fig1}(e). In a recent study, using the technique described in Fig.\;\ref{fig:Fig1}, we indeed measured the Fermi contour anisotropy of 2D hole systems confined to a GaAs quantum well and found that the contour is severely distorted when the 2D holes are subjected to a strong $B_{||}$ of the order of 10 T \cite{Kamburov.2012b}. In the work described here, we use similar samples and techniques to demonstrate that the Fermi contour of the hole-flux CFs is also distorted when a strong $B_{||}$ is applied, although the degree of anisotropy is much smaller.

We studied strain-induced superlattice samples with lattice periods of $a=175$ and 200 nm from a 2D hole system confined to a 175-\AA\-wide GaAs quantum well grown via molecular beam epitaxy on a (001) GaAs substrate. The quantum well, located 131 nm under the surface, is flanked on each side by 95-nm-thick Al$_{0.24}$Ga$_{0.76}$As spacer layers and C $\delta$-doped layers. The 2D hole density at $T\simeq$ 0.3 K is $p\simeq 1.5 \times 10^{11}$ cm$^{-2}$, and the mobility is $\mu=1.2\times10^{6}$ cm$^2$/Vs. As schematically illustrated in Fig.\;\ref{fig:Fig1}(e), the sample has two Hall bars, oriented along the [$110$] and [$\overline{1}10$] directions. The Hall bars are covered with periodic gratings of negative electron-beam resist. Through the piezoelectric effect in GaAs, the resist pattern induces a periodic density modulation \cite{Skuras.APL.1997,Endo.PRB.2000,Endo.PRB.2001,Endo.PRB.2005,Kamburov.PRB.2012b, Kamburov.2012b, Kamburov.PRL.2012}. We passed current along the two Hall bar arms and measured the longitudinal resistances along the arms in tilted magnetic fields, with $\theta$ denoting the angle between the field direction and the normal to the 2D plane; see Fig.\;\ref{fig:Fig1}(e). The sample was tilted around the [$\overline{1}10$] direction so that $B_{||}$ was always along [$110$]. We performed the experiments using low-frequency lock-in techniques in two $^3$He refrigerators with base temperatures of $T\simeq$ 0.3 K, one with an 18 T superconducting magnet and the other with a 31 T resistive magnet.

\begin{figure}[t]
\includegraphics[trim=0.6cm 0.3cm 0cm 0.5cm, clip=true, width=0.45\textwidth]{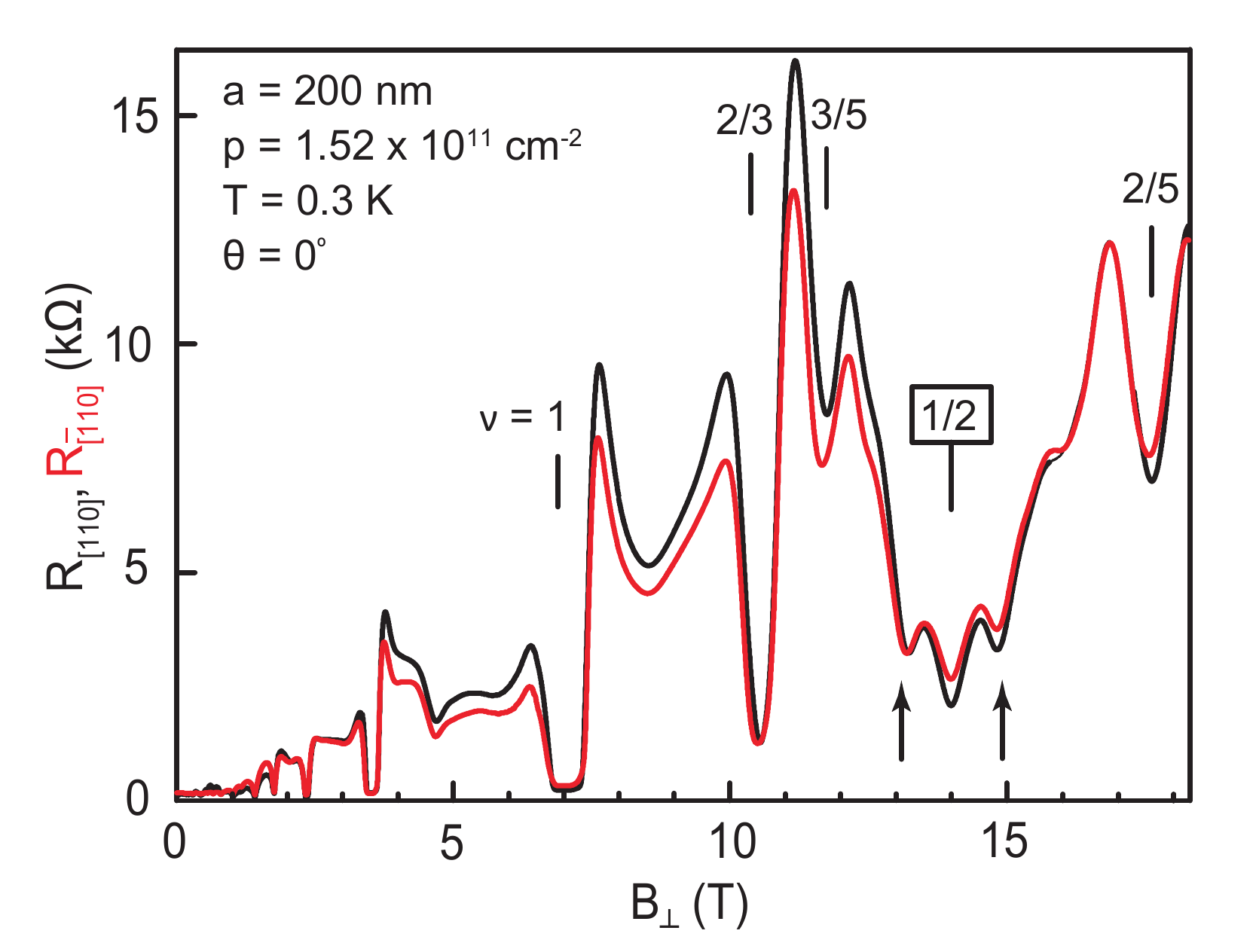}
\caption{\label{fig:Fig2} (color online) Magnetoresistance traces from the [$110$] and [$\overline{1}10$] Hall bars of a sample with $a=200$ nm are shown in black and red, respectively. The two prominent resistance minima visible near $\nu=1/2$, marked by arrows, signal the commensurability of the CF cyclotron orbit diameter with the period of the density modulation (see text).}
\end{figure}

\begin{figure}[t]
\includegraphics[trim=0.1cm 0.6cm 0.0cm 0.4cm, clip=true, width=0.45\textwidth]{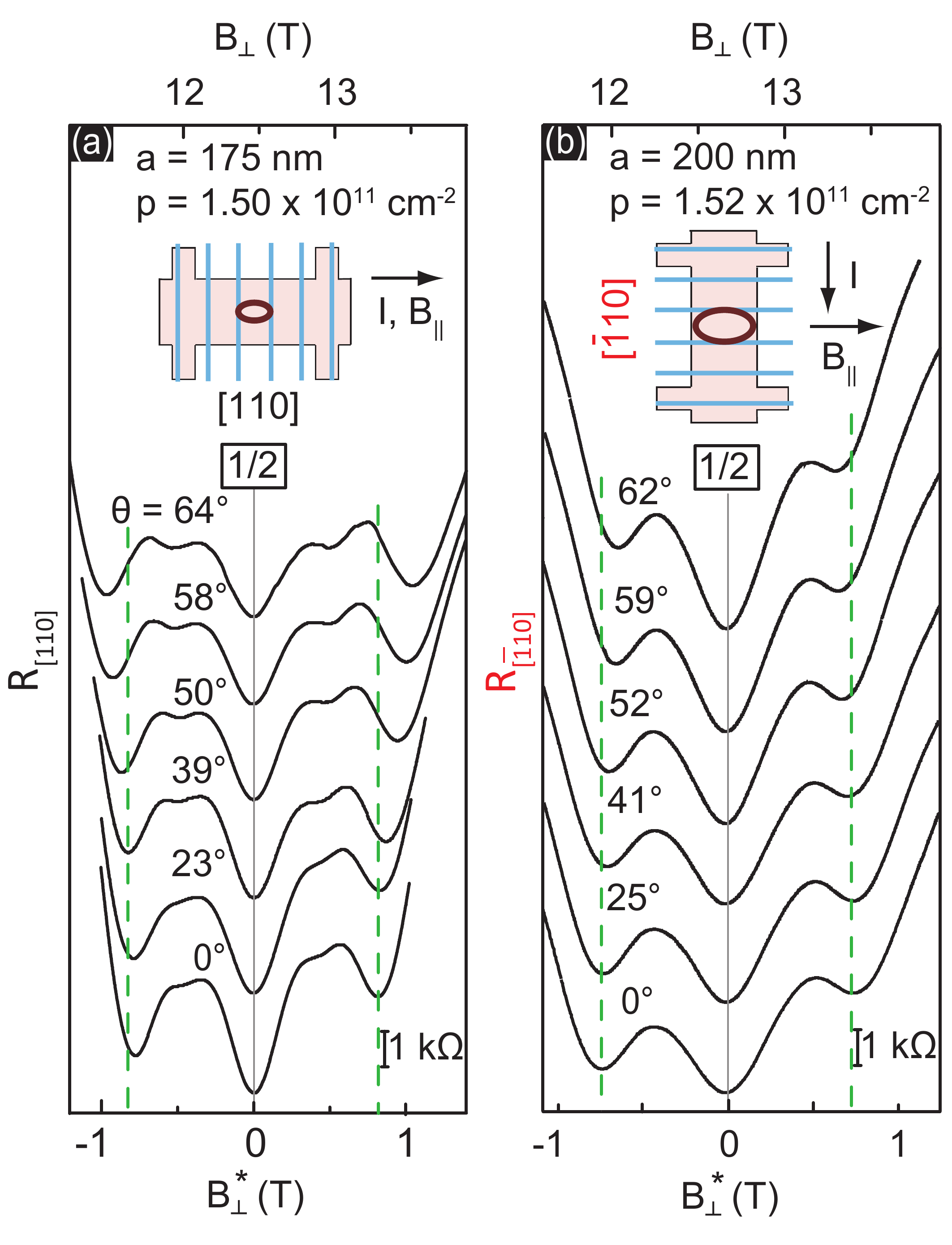}
\caption{\label{fig:Fig3} (color online) (a) Summary of the evolution of the magnetoresistance in the vicinity of $\nu=1/2$ of the $a=175$ nm sample measured along the [$110$] Hall bar. The tilt angle $\theta$ is given for each trace. The vertical green dashed lines mark the expected positions of the primary CF commensurability resistance minima if the CF cyclotron orbit were circular. (b) Magnetoresistance data for the $a=200$ nm sample measured along the [$\overline{1}10$] Hall bar. In both (a) and (b), the scale for the applied \textit{external} field $B_{\perp}$ is shown on top while the scale for the \textit{effective} magnetic field $B^*_{\perp}=B_{\perp}-B_{\perp,1/2}$ felt by the CFs is given at the bottom ($B_{\perp,1/2}$ is the external field at $\nu=1/2$).}
\end{figure}

The high-field data (Fig.\;\ref{fig:Fig2}), taken at $\theta=0^{\circ}$ ($B_{||}=0$), for the two Hall bars of the $a=200$ nm sample exhibit prominent commensurability features around $\nu=1/2$: a characteristic, V-shaped, resistance dip centered at $\nu=1/2$ and two strong resistance minima, marked by arrows on each side of $\nu=1/2$, followed by flanks of rapidly rising resistance \cite{Willett.PRL.1997,Smet.PRB.1997,Smet.PRL.1998,Mirlin.PRL.1998,Oppen.PRL.1998,Smet.PRL.1999,Zwerschke.PRL.1999,Kamburov.PRL.2012}. Of particular interest to us are the two minima as they correspond to the commensurability of the CF cyclotron orbit diameter ($2R_C^{*}$) with the period $a$ of the potential modulation. Quantitatively, for a circular CF Fermi contour, the positions of these resistance minima are given by the magnetic commensurability condition \cite{Willett.PRL.1997,Smet.PRB.1997,Smet.PRL.1998,Mirlin.PRL.1998,Oppen.PRL.1998,Smet.PRL.1999,Zwerschke.PRL.1999,Kamburov.PRL.2012,footnote103}:
\begin{equation} \frac{2R_C^*}{a}=\frac{5}{4},
\label{eq:1}
\end{equation}
where $R_C^*=\hbar k_F^*/eB_{\perp}^*$ is the CF cyclotron radius at $B_{\perp}^*$, $k_F^*=\sqrt{4\pi p}$ is the CF Fermi wave vector, and $p$ is the 2D hole density \cite{footnote100}; note that the expression for $k_F^*$ takes into account complete spin polarization at high fields and is larger that its low-field hole counterpart by a factor of $\sqrt{2}$. In a recent study, it was demonstrated that Eq.\;(\ref{eq:1}) indeed describes the positions of resistance minima exhibited by hole-flux CFs in our samples in the absence of $B_{||}$ \cite{Kamburov.PRL.2012}; this is seen in Fig.\;\ref{fig:Fig2} where the arrows point to the positions of the minima expected from Eq.\;(\ref{eq:1}). In the present study, we monitor the shift in the observed positions of these minima as a function of applied $B_{||}$ to directly probe the size and shape of the CF Fermi contour.

\begin{figure*}[t]
\includegraphics[trim=0.6cm 0.3cm 0cm 0cm, clip=true, width=0.89\textwidth]{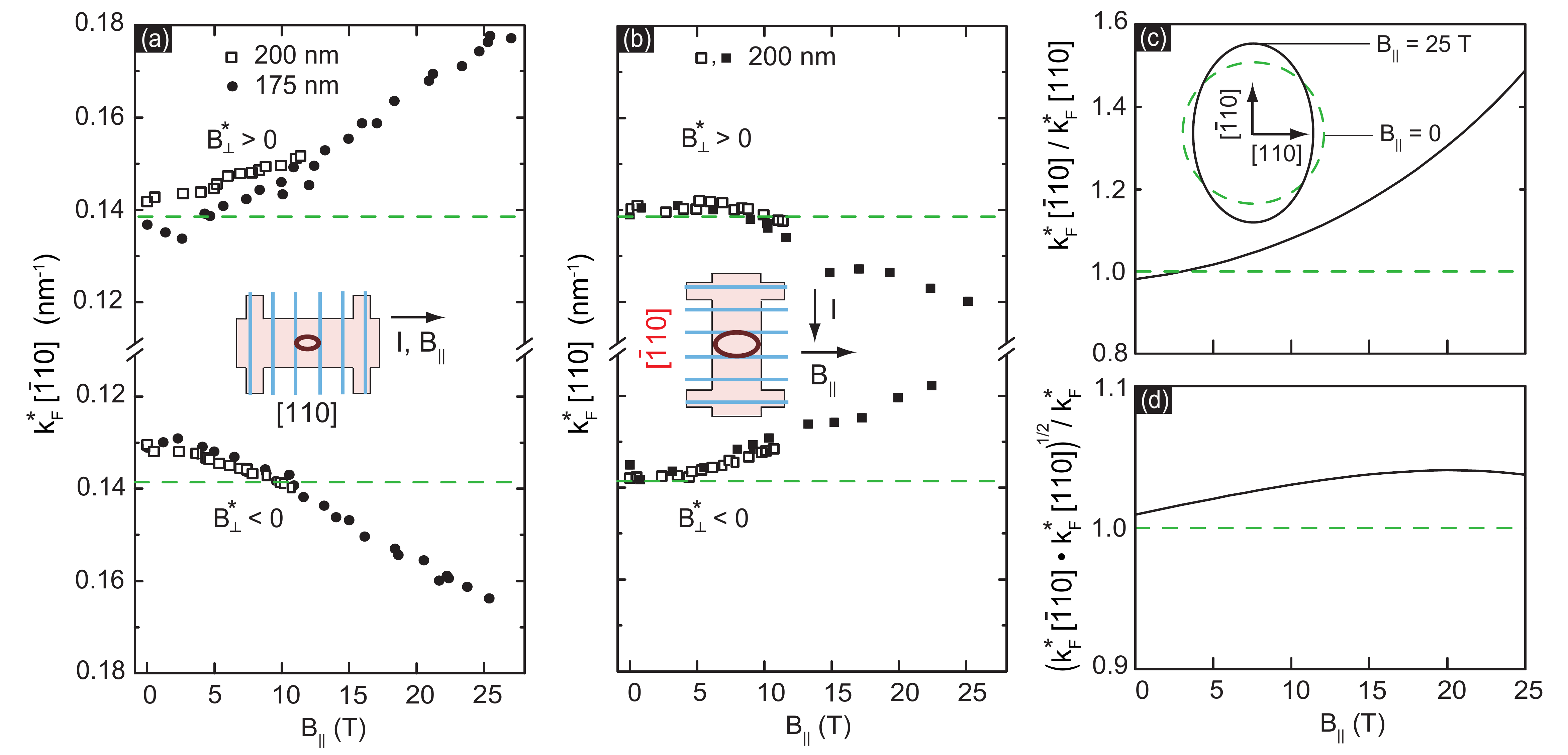}
\caption{\label{fig:Fig4} (color online) (a), (b) Measured values of the CF Fermi wave vectors $k_F^*$ along the [$\overline{1}10$] and [$110$] directions, respectively. Data shown with open squares were measured in a superconducting magnet with maximum field of 18 T. The closed symbols are data taken in a 31 T resistive magnet system. Horizontal green dashed lines represent the Fermi wave vector $k_F^*=\sqrt{4 \pi p}$, expected for a circular CF Fermi contour (see Eq.\;(\ref{eq:1})). The 2D hole density is $p\simeq 1.5 \times 10^{11}$ cm$^{-2}$. (c) Relative anisotropy of the CF Fermi contour deduced from dividing the (interpolated) measured values of $k_F^*$ along [$\overline{1}10$] by those along [$110$]; data for $B^*_{\perp}>0$ were used. The inset schematically shows the CF Fermi contour at $B_{||}=0$ (dashed, green circle) and at $B_{||}=25$ T (solid, black curve); the latter is based on the assumption that the CF is elliptical. (d) The geometric mean of the measured $k^*_F$ along [$\overline{1}10$] and [$110$], divided by $k^*_F$ expected for a circular CF Fermi contour.  }
\end{figure*}

As illustrated in Fig.\;\ref{fig:Fig3}, the application of $B_{||}$ has a profound effect on the appearance of the commensurability minima near $\nu=1/2$. Data for the two Hall bars along the [$110$] and [$\overline{1}10$] directions are shown side-by-side in Figs.\;\ref{fig:Fig3}(a) and (b). In both panels, the vertical green dashed lines mark the expected positions of the CF commensurability resistance minima based on Eq.\;(\ref{eq:1}). These dashed lines match very well the observed positions of the resistance minima for the bottom traces of  Fig.\;\ref{fig:Fig3} which were taken at $\theta=0$ ($B_{||}=0$) \cite{footnote101}. With increasing $\theta$ and $B_{||}$, for the [$110$] Hall bar (Fig.\;\ref{fig:Fig3}(a)), the positions of the two resistance minima shift away from the dashed lines to \textit{higher} values of $|B^*_{\perp}|$. As evidenced by the top trace in Fig.\;\ref{fig:Fig3}(a), their shift reaches $\simeq 0.25$ T at the highest $\theta$ ($=64^o$). In contrast, the positions of the resistance minima for the Hall bar in the perpendicular, [$\overline{1}10$] direction (Fig.\;\ref{fig:Fig3}(b)) move towards \textit{lower} $|B_{\perp}^*|$, and the shift is smaller. In particular, when $\theta=62^o$, the minima of the top trace have moved toward $B_{\perp}^*=0$ only by $\simeq 0.10$ T.

The positions of the resistance minima along the [$\overline{1}10$] and [110] directions can be used to directly extract the magnitude of the CF Fermi wave vectors along [110] and [$\overline{1}10$] and [110] respectively. According to Eq.\;(\ref{eq:1}), $k^*_F=(5/8)(eaB^*_{\perp}/\hbar)$, where $B^*_{\perp}$ indicates the effective CF magnetic field at which the resistance minimum is observed. Note that the commensurability condition along a given modulation direction gives the size of $k^*_F$ in the direction \textit{perpendicular} to the modulation direction \cite{Kamburov.2012b,Gunawan.PRL.2004,Kamburov.PRB.2012b}. Using the above relation, we converted the $B_{\perp}^*$ positions of the resistivity minima seen in Fig.\;\ref{fig:Fig3} to the size of the CF $k^*_F$ along the [$\overline{1}10$] and [110] directions and summarize the results in Figs.\;\ref{fig:Fig4}(a) and (b). The horizontal, green dashed lines in these figures indicate the expected $k^*_F$, if a circular CF Fermi contour is assumed. For data taken at $B_{||}=0$, the values of $k^*_F$ are mostly in good agreement with those expected for CFs with circular Fermi contour
\cite{footnote101}. With increasing $B_{||}$, however, it is clear in Figs.\;\ref{fig:Fig4}(a) and (b) that the CF Fermi wave vector along [$\overline{1}10$] increases (by as much as 30\% at the highest $B_{||}$), while along [$110$] it decreases (by nearly 15\%). These data therefore provide unambiguous and quantitative evidence for a deformation of the CF Fermi contour in the presence of an applied $B_{||}$. Moreover, by tilting the sample, the CF anisotropy can be controllably tuned.

We combine the data of Figs.\;\ref{fig:Fig4}(a) and (b) to deduce the relative distortion of the CF Fermi contour, as shown in Fig.\;\ref{fig:Fig4}(c). Here we plot the ratio of $k^*_F$ along the [$\overline{1}10$] and [$110$] directions, as deduced from the $B^*_{\perp} > 0$ dada. Before performing the division, we fitted each set of data points from Fig.\;\ref{fig:Fig4}(a) and (b) with simple, second-order polynomials. The ratio of the two $k^*_F$ values in the two directions is as high as 50\% at $B_{||}=25$ T, indicating a severe distortion as a result of $B_{||}$. One obvious question that arises is whether the CF Fermi contour is elliptical or has a more complicated (warped) shape, e.g., similar to those we recently measured for 2D holes near zero magnetic field \cite{Kamburov.2012b}. Since in our experiments we measure CF $k^*_F$ only along two specific (and perpendicular) directions, we cannot rule out a complicated shape. However, our data are consistent with a nearly elliptical CF Fermi contour. This is evinced from the plot of Fig.\;\ref{fig:Fig4}(d) where we plot the ratio of the geometric mean of the two $k^*_F$'s we measure along [$\overline{1}10$] and [$110$] to the Fermi wave vector expected for a circular CF Fermi contour, i.e., to $k^*_F=\sqrt{4 \pi p}$. The fact that this ratio is close to unity implies that the area enclosed by an elliptical Femri contour whose major and minor Fermi wave vectors are equal to the two values we measure has the correct magnitude, i.e., it accounts for all the CFs. We show such an ellipse in Fig.\;\ref{fig:Fig4}(c) inset (solid black curve)

The CF commensurability data described here provide the first direct evidence that the CF Fermi contour can be anisotropic. Moreover, they demonstrate how this anisotropy can be tuned via the application of a strong $B_{||}$. The origin of this anisotropy is very likely the coupling between $B_{||}$ and the out-of-plane motion of the CFs, which have non-zero thickness. Such coupling is known to severely distort the Fermi contour of \textit{low-field} carriers (Fig. 1). Indeed, for the low-field 2D holes in our samples, we recently measured a very elongated (and non-elliptical) Fermi contour with an anisotropy ratio of about 3 (at $B_{||}=~15$ T), and found the data to be in good agreement with the results of band calculations \cite{Kamburov.2012b}. The anisotropy ratio we measure for the CF Fermi contour at a comparable $B_{||}$ is much smaller, only about 1.2 (Fig.\;\ref{fig:Fig4}(c)). Absent, however, are theoretical calculations that would treat the anisotropy of CF Fermi contours in the presence of $B_{||}$ in general, and in particular explain the anisotropy we measure in our experiments. The much different anisotropy that we observe for the hole-flux CF Fermi contour compared to the 2D holes indeed appears to contradict the conclusions of the only available theoretical work which predicts that the CF Fermi contour shape should be identical to that of the zero-field particles \cite{Balagurov.PRB.2000}. We note that, besides its thickness, other parameters of the quasi-2D carrier system, such as the band structure and effective mass as well as the character of the Landau level where the CFs are formed, are also likely to play an important role in determining the anisotropy of the CF Fermi contour in a strong $B_{||}$. Our conjecture is based on our preliminary data for a 300-\AA\-wide GaAs quantum well containing \textit{electrons}: despite its larger thickness, this sample exhibits a CF Fermi contour anisotropy which is much smaller than the anisotropy we observe in our 175-\AA\-wide GaAs \textit{hole} quantum well sample.

While a quantitative explanation of our experimental data awaits future theoretical work, we emphasize that our results clearly establish the presence of CF Fermi contour anisotropy. This has important implications and raises several interesting questions. For example, what is the role of anisotropic interaction in general? How does the anisotropy affect the ground states and the excitations of the 2D carrier system at high perpendicular fields? Does the anisotropy affect, e.g., the energy gaps of the fractional quantum Hall states?  Our results provide stimulus for future studies to answer some of these questions.

\begin{acknowledgments}
We acknowledge support through the DOE BES (DE-FG02-00-ER45841) for measurements, and the Moore and Keck Foundations and the NSF (ECCS-1001719, DMR-0904117, and MRSEC DMR-0819860) for sample fabrication and characterization. This work was performed at the National High Magnetic Field Laboratory, which is supported by NSF Cooperative Agreement No. DMR-0654118, by the State of Florida, and by the DOE. We thank J.K. Jain and R. Winkler for illuminating discussions, and S. Hannahs, T. Murphy, and A. Suslov at NHMFL for valuable help during the measurements. We also thank Tokoyama Corporation for supplying the negative electron-beam resist TEBN-1 used to make the samples.
\end{acknowledgments}


\begin{thebibliography}{99}
\bibitem{footnote100} Throught this paper, we use "*" to denote the CF parameters.
\end{thebibliography}
\end{document}